\newcommand{\kms}{{~\rm km\; s^{-1}}}

\newcommand{\msyr}{{M_{\odot}~\rm yr^{-1}}}
\newcommand{\cm}{{~\rm cm}}
\newcommand{\s}{{~\rm s}}
\newcommand{\km}{{~\rm km}}

\newcommand{\K}{{~\rm K}}
\newcommand{\erg}{{~\rm erg}}
\newcommand{\yr}{{~\rm yr}}

\newcommand{\Mpc}{{~\rm Mpc}}

\newcommand{\AU}{{~\rm AU}}

\newcommand{\days}{{~\rm days}}

\documentclass[12pt,preprint,a4paper]{aastex}

\shorttitle{TRANSIENT}

\usepackage{amsmath}                
\usepackage{amsfonts}               
\usepackage{amssymb}                
\usepackage{epsfig}                 

\begin{document}

\title{NGC~300~OT2008-1 AS A SCALED-DOWN VERSION OF THE ETA CARINAE GREAT ERUPTION}

\author{Amit Kashi\altaffilmark{1}, Adam Frankowski\altaffilmark{1}, and Noam Soker\altaffilmark{1}}

\altaffiltext{1}{Department of Physics,
Technion$-$Israel Institute of Technology, Haifa 32000, Israel;
kashia@physics.technion.ac.il; adamf@physics.technion.ac.il; soker@physics.technion.ac.il.}

\begin{abstract}
We propose that the intermediate luminosity optical transient
NGC~300~OT2008-1 was powered by a mass transfer episode from an
extreme Asymptotic Giant Branch star to a Main Sequence companion.
We find a remarkable similarity of the shapes of the light curves of
the several months long NGC~300~OT2008-1 outburst,
of the three months long 2002 enigmatic outburst of the B star V838~Mon,
and the twenty-years long Great Eruption of the massive binary system Eta Carinae
that occurred in the 19th century.
Their similar decline properties hint to a common energy source:
a gravitational energy that is released by accretion onto a main sequence star.
These events populate a specific strip in the total energy vs.
outburst duration diagram.
The strip is located between novae and supernovae.
We add recent transient events to that diagram and find them to occupy
the same strip.
This suggests that some intermediate luminosity
optical transients are powered by accretion onto a compact object
(not necessarily a main sequence star).
These transients are expected to produce bipolar ejecta
as a result of the geometry of the accretion process.
\end{abstract}


\section{INTRODUCTION}
\label{sec:introduction}

The Intermediate Luminosity Optical Transient (ILOT) NGC~300~OT2008-1 (hereafter NGC~300~OT)
was discovered in Apr 24, 2008 (Monard 2008), a few weeks after its assumed outburst.
The distance to the host galaxy NGC~300 is $\sim 1.88 \Mpc$ (Gieren et al. 2005; Freedman et al. 2001).
With a bolometric luminosity of $L_{\rm bol} = 1.6 \times 10^{40} \erg
\s^{-1}$ at discovery (Bond et al. 2009), it is located between
supernovae (SNe) and novae, and cannot be easily grouped within any
subtype of these two classes (Kulkarni \& Kasliwal 2009).
The pre-outburst progenitor system was discovered by Prieto (2008).
It was enshrouded by dust (Bond et al. 2009; Berger et al. 2009;),
and had a luminosity of about $6\times10^4 L_\odot$, corresponding to a
$M = 10 - 15 M_\odot$ star, probably in the extreme Asymptotic Giant Branch (AGB) stage (Thompson et al. 2008).
A more massive red supergiant of mass $M = 12 - 25 M_\odot$, as found by
Gogarten et al. (2009) based on stellar evolution considerations, would also be consistent with the data.

Berger et al. (2009) could fit the UV-visual spectral energy distribution (SED)
during outburst with a blackbody spectrum with a temperature of $T_B = 4670 \pm 140 \K$,
and a radius of $R_B = 14.7 \pm 1.4 \AU$.
But if inter- and circumstellar extinction is considered (Bond et al. 2009),
based on the appearance of the absorption spectrum at maximum,
a temperature of $\sim 7500\K$ is more consistent with the SED.
The source was not detected neither in the X-ray band
($F_x < 1.2 \times 10^{-14} \erg \s^{-1} \cm^{-2}$) nor in the radio band.
Berger et al. (2009) attributed the $\sim 10^3 \kms$ red wing of the
Ca II H\&K absorption lines either to an infalling gas from a previous
eruption or to a wind of a companion star.
In either case the star accreting this matter is likely to be a main sequence (MS) star.
Together with the evidence for the supergiant nature of the progenitor, this
implies that there are two different stars in the system.
Bond et al. (2009) interpreted the Hydrogen Balmer lines' and the Ca II IR triplet's double features
as indicating the presence of a bipolar outflow expanding at a velocity of $\sim 75 \kms$.
None of the observed lines exhibited P-Cygni profiles or velocities
exceeding $10^3 \kms$ (Berger et al 2009); the latter property is a clear
departure from typical SN behavior.

According to Prieto et al. (2009) the ILOT event that most resembles
NGC~300~OT is SN~2008S (Prieto et al. 2008; Wesson et al. 2009).
Both ILOTs were the result of an energetic eruption in a dust-enshrouded
$10-20 M_\odot$ star that survived the eruption (namely,
SN~2008S was not a SN at all).
Smith et al. (2009) suggested that the physical mechanism which produced
SN~2008S is a super-Eddington wind, similar to the super-outbursts of massive
Luminous Blue Variables (LBVs).
Botticella et al. (2009) suggested that the progenitor was an extreme (``super'') AGB star.
Bond et al. (2009) suggested that both SN 2008S and NGC~300~OT originated form evolved massive stars
on a blue loop to warmer temperatures, and were subjected to increased instability due to prior mass loss.

An asymmetric dusty environment extending a few-thousand$\AU$
surrounding NGC~300~OT (Patat et al. 2009)
hints to a previous possible eruption.
This asymmetry may further hint to the presence of a companion star,
although we note that up to date there has been no definite observation proving
{{{}}} a companion {{{existence}}}.
Thompson et al. (2008) suggested that {{{}}} ILOTs occur due to single
star processes, e.g., electron-capture SN, an explosive birth of a massive WD,
or an enormous outburst of a massive star.
In {{{their model for}}}
NGC~300~OT and SN~2008S the progenitors were luminous ($\sim 4 - 6
\times 10^4 L_\odot$) dust-enshrouded stars, at the end of their AGB stage.

Based on the model proposed by Soker (2004), we examine in
this \emph{Letter} (Section \ref{sec:companion}) whether an eruptive mass transfer
{{{onto}} a companion can account for the {{{}}}properties of NGC~300~OT.
In that model most of the energy of the outburst was gravitational energy
released by $\sim {\rm few} \times 0.1 M_\odot$ accreted by a MS B-type companion.
The binary system survives the event.
In Section \ref{sec:transients} we suggest a physical mechanism to account for most
of the gravitationally-powered outbursts.

\section{ACCRETION BY A SURVIVING COMPANION}
\label{sec:companion}

We examine a model where the source of the mass is an extreme-AGB star, while the main
source of the energy is a gravitational energy released by the mass accreted
onto a MS companion.

Based on previous papers (Berger et al. 2009; Bond et al. 2009; Gogarten et al. 2009; Prieto et al. 2009)
we scale the mass of the extreme-AGB star by $M_1 \simeq 15M_\odot$.
We also assume that the main sequence companion is not much lighter than the
primary, and scale it with $M_2 \simeq 8 M_\odot$, corresponding to a MS radius
of $R_2 \simeq 3.5 R_\odot$ (our model can work for $3 M_\odot  \la M_2 \la 10 M_\odot$).
We attribute the slow outflow of $\sim 50-75 \kms$ that probably contains most of the mass
to the extreme-AGB star, as it is about equal to its escape speed
(assuming a radius of $R_1 \simeq 5 \AU$).
The faster outflow of up to $\sim 600 \kms$ fits better the escape velocity from the
MS companion.

Our scenario starts with some kind of instability in the extreme-AGB star that causes
a mass of several$\times 0.1 M_\odot$ to be lost from the star at a slow velocity.
In our model the instability is not the source of the extra energy.
Therefore, we can assume that the instability does not increase much, or even reduces,
the primary luminosity.
We do not specify the source of the instability, but it might be, for example,
{{{}}}a strong magnetic eruption.
AGB stars are known to have extensive convective region with strong convection
(the convective cells have a relative high velocity and long mixing length).
If the star has a non-negligible rotation due to its tidal interaction with the companion,
then a strong magnetic activity might be expected {{{{(e.g, Garcia-Segura et al. 2001)}}}}.
The magnetic eruption causes the primary to overfill its Roche lobe, resulting in
both mass loss from the system and mass transfer to the companion.

According to Bond et al. (2009), the transient at maximum light exhibited a spectrum
of an F superigant.
The underlying radiation source {\em during} the event could be even hotter, if the observed spectrum
was produced in an optically thick wind.
However, this cannot be used to constrain the spectral type of the progenitor
{\em before} the event.
We take the most conservative approach, and take the star to have the lowest temperature possible
for its mass at this evolutionary stage (extreme-AGB), $\sim 3500 \K$.
For an extreme AGB effective temperature of $\sim 3500 \K$ the radius of the
progenitor of NGC~300~OT was $R_1 \sim 3 \AU$.
The dynamical time scale for a mass of $M_1=15 M_\odot$ is $\sim 5~$months.
The outburst duration of $80 \days$ can be understood as dynamical time scale.
Namely, in our model a dynamical instability lead to high mass transfer episode.
The super-Eddington luminosity of the accreting secondary might have
helped in terminating the high mass transfer rate.
For an efficient accretion the companion in our model of NGC~300~OT has to be
very close to the primary, $\sim 2 R_1$.
We note that if the orbit is highly eccentric, say $e=0.9$ as in $\eta$ Car,
and periastron passage is at $a_p=2R_1 \simeq 6 \AU$, then the orbital period is
$\sim 100 \yr$.
Furthermore, if the outburst was caused by the periastron passage, then
some high mass loss episode could have occurred 100 years ago, but not necessarily as strong.
Assuming a velocity of $75 \km \s^{-1}$ (as observed for the present bipolar outflow)
these ejecta are at a distance of $\sim 1500 AU$.
If the progenitor was indeed an F star{{{}}}, its higher temperature
would imply a smaller radius, and hence a shorter dynamical timescale of about half a month
for an effective temperature of $ 7500 \K$.
This will considerably ease the constraints on our model, as the outburst in our model is
limited from below by the dynamical time scale.

In our scenario most of the outburst energy comes from accretion onto the companion.
Therefore, we can use the observed luminosity and total energy to constrain the accretion
mass and time scale.
The total radiated energy is $E_{\rm{rad}}=10^{47} \erg$ (Bond et al. 2009; Berger et al. 2009).
Adding the kinetic energy of the ejected mass doubles (or even more) the total
eruption energy, which we scale with $E_{\rm{tot}} = 2 \times 10^{47} \erg$.
The accreted mass onto the companion is constrained to be
\begin{equation}
\begin{split}
\Delta M_{\rm acc} = 0.05
\left( \frac{E_{\rm{tot}}}{2 \times 10^{47} \erg} \right)
\left(\frac{M_2}{8 M_\odot}\right)^{-1} \\
\times \left(\frac{R_2}{3.5 R_\odot}\right)
\left(\frac{\chi}{0.5}\right)^{-1}
M_\odot,
\label{eq:deltaM}
\end{split}
\end{equation}
where $\chi$ is the efficiency of converting gravitational energy to eruptive energy.
If the accreting star does not rotate fast, then $\chi$ is very close to 1.
Here we take a conservative approach, and scale with $\chi=0.5$.

The time scale of the high state of the eruption was $\sim 60 \days$ (Bond et al. 2008).
This would give an average mass accretion rate of $\gtrsim 0.3 M_\odot \yr^{-1}$,
depending on $\chi$ and the other parameters.

There are some similarities between our suggested model of NGC~300~OT  and
the merger model of the outburst of V838 Mon (Tylenda \& Soker 2006).
In both ILOTs the energy is obtained from a process of mass transfer.
In V838 Mon, according to the model proposed by Tylenda \& Soker (2006),
it was a merger process, in which a $\sim 0.3 M_\odot$ star
merged with a more massive one, on a timescale of $\sim 80 \days$.
This results in an accretion rate of $\sim 1.4 \msyr$, close to that of NGC~300~OT.
The ejection velocities in V838 Mon ($\sim 300 \kms$) are also similar.
The duration of the V838 Mon eruption is also quite similar to that of NGC~300~OT.
Indeed, the total energy of the two ILOTs is few~$\times 10^{47} \erg${{{}}}.

In some sense the NGC~300~OT eruption is a scaled-down version of the
Great Eruption (GE) of $\eta$ Car.
The GE of $\eta$ Car in 1843 ejected more than $10 M_\odot$
(Smith et al. 2003) from the $100-150 M_\odot$ progenitor LBV star.
Soker (2004) suggested that a large fraction of this mass was accreted
by the secondary star, of mass $\sim 25 M_\odot$.
Part of the mass transferred to the companion was expelled in a bipolar outflow,
that shaped the bipolar structure of $\eta$ Car, the Homunculus
(e.g., recent review by Smith 2009).
In {{{$\eta$ Car}}} more data are available from observations,
e.g. we know the binary period and the ejecta energy,
therefore the accretion-powered model could be worked out quantitatively
in more detail (Soker 2004).

It is possible that the trigger for the impulsive mass loss episode in $\eta$ Car
and NGC~300~OT was the same.
Harpaz \& Soker (2009) suggested that the mass loss from the primary in the
GE of $\eta$ Car was triggered by magnetic eruption in the primary envelope.
The magnetic eruption does not increase much the luminosity, but mainly causes an impulsive
mass loss episode from the primary.
The secondary accretes a large fraction of the outflowing mass.

We now turn to discuss the light curves (Fig. \ref{fig:lightcurves1}).
The early light curve of {{{V838~Mon}}} showed 3 peaks (Munari et al. 2002),
and then had a rapid decay by 5 magnitudes in about 20 days,
shorter than the $\sim 80 \days$ duration of the high-brightness phase.
It is unknown if such peaks existed in NGC~300~OT (Bond et al. 2009),
but this may be due to a poor time coverage of the early light curve.
However, in the decline phase, the light curve of NGC~300~OT resembles those of V838 Mon
and $\eta$ Car GE (Fig. \ref{fig:lightcurves1}).
The similarity of NGC~300~OT to V838~Mon and few other ILOTs was already noticed
by other authors (e.g. Berger et al. 2009, Smith et al. 2009).
\begin{figure}[!t]
\resizebox{0.89\textwidth}{!}{\includegraphics{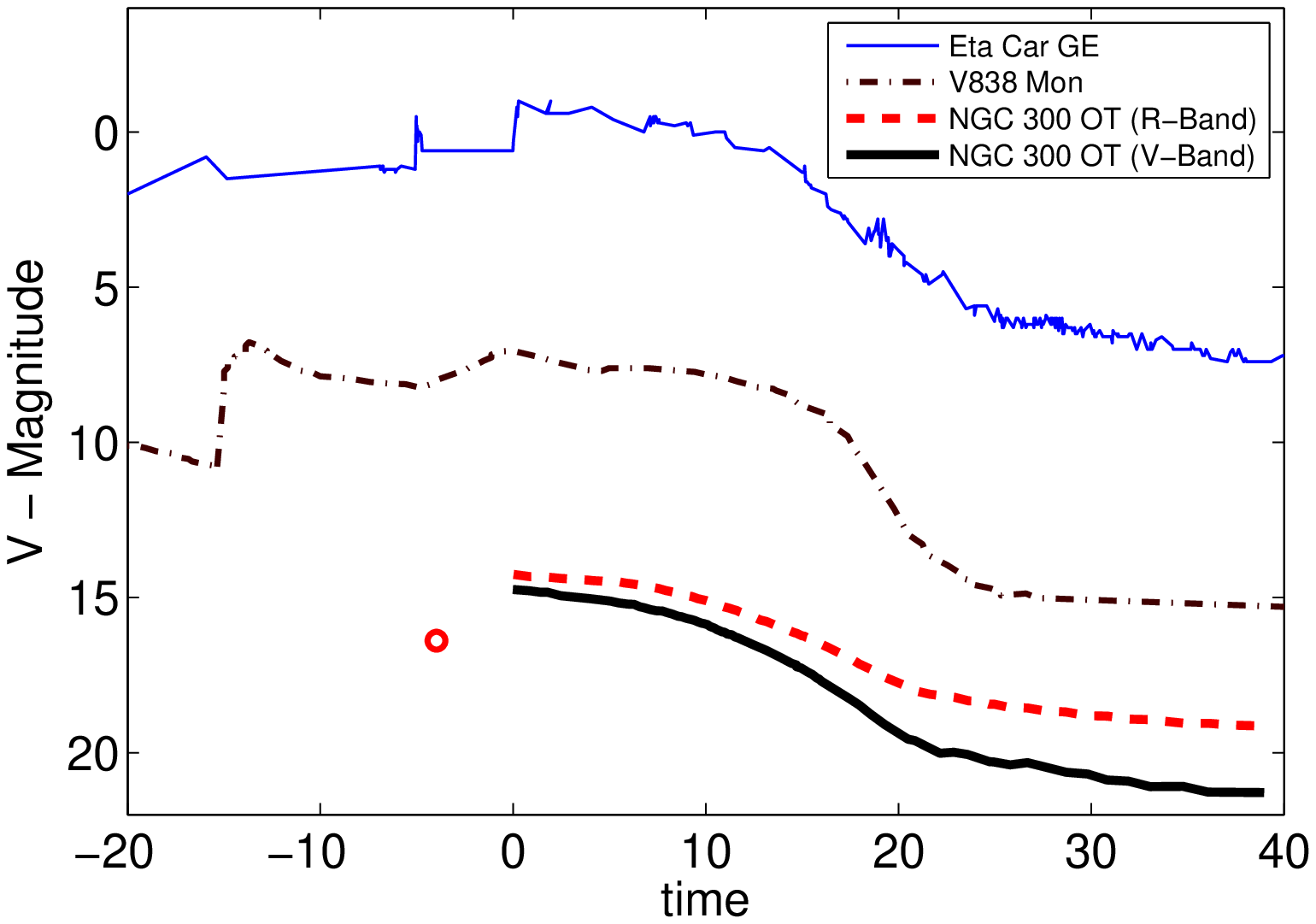}}
\resizebox{0.89\textwidth}{!}{\includegraphics{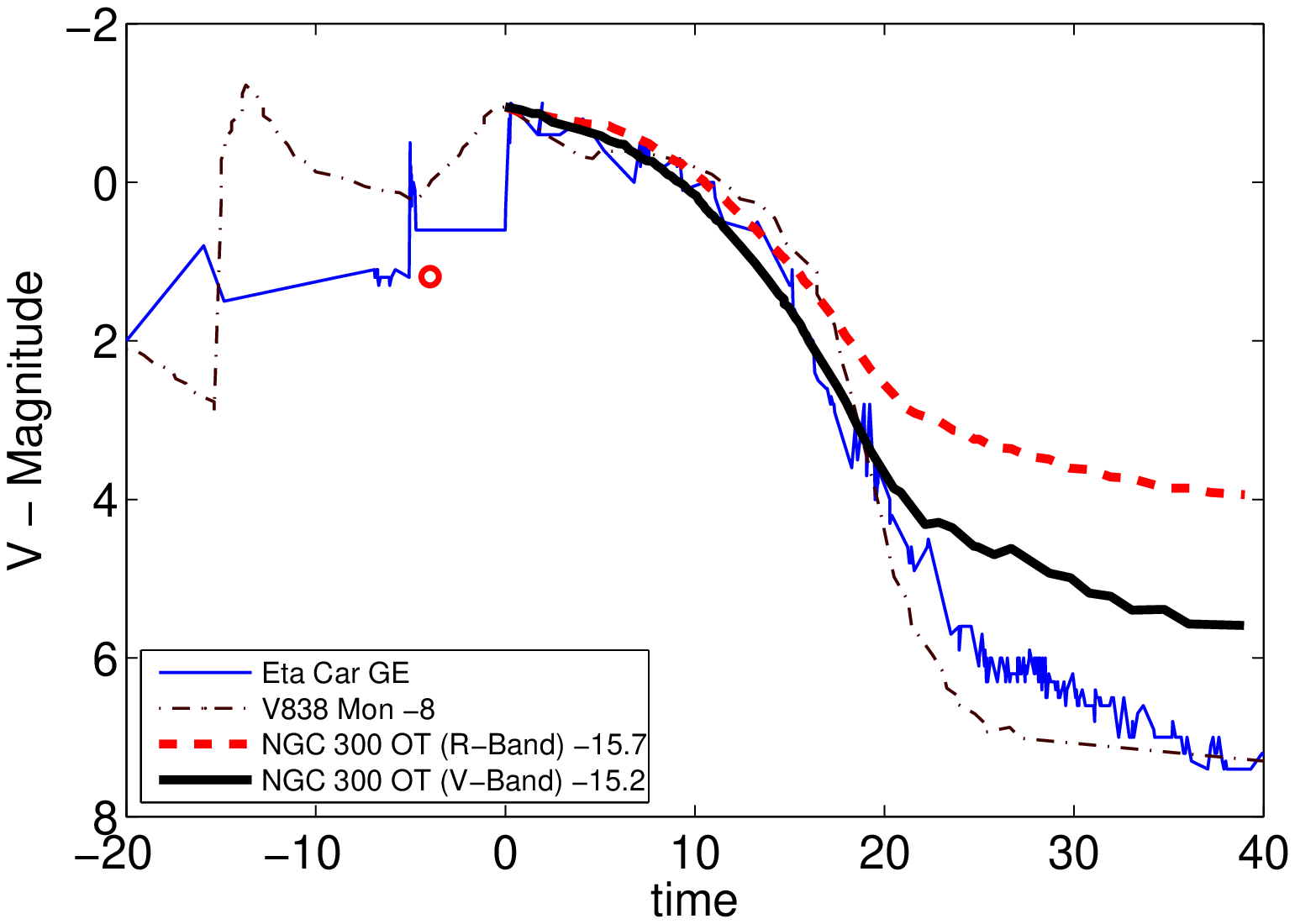}}
\caption{\footnotesize{Comparison of the V-band light curves of the $\eta$ Car Great Eruption (GE), V838 Mon and NGC~300~OT.
The timescale was normalized so that
1 time unit equals 1 year for $\eta$ Car GE, 2.2 days for V838 Mon, and 5.6 days for NGC~300~OT.
For NGC~300~OT the R-band is also plotted, for which there has been one observation before the maximum (Bond et al. 2009),
marked with a red circle.
Top: the three separated light curves; the apparent V-mag axis was not rescaled.
Bottom: The same curves translated vertically to bring peak luminosities to overlap (see legend for the shift values).
It can be easily seen that the slope of the decline phase and its rate of change are
similar for the three eruptions.
}}
\label{fig:lightcurves1}
\end{figure}

In the case of NGC~300~OT, 120 days were required for a decay by 5 magnitudes,
a time longer than the high-brightness phase.
For the GE of $\eta$ Car, it took no less than $\sim 18 \yr$ for the light curve
to decay by 5 magnitudes (Humphreys et al. 1999).

Remarkably, when we normalize the timescales of the three objects without changing the magnitude scale
we find that the light curves look very similar (Fig. \ref{fig:lightcurves2}).
A similar procedure is known as ``stretch correction'' when applied to SN light curves.
The normalized slope of the decline behaves the same way for all three eruptions.
We consider this a hint that all three eruptions are governed by the
same physical mechanism.
For example, the residual energy source is not as high
(even when the bolometric light curve is considered)
as in radioactive decay in SNe.
Weaker contribution is possible, e.g. from mass accretion at a low rate (like in the
scenario for V838 Mon; Tylenda 2005).
Our leading candidate for the powering mechanism is dissipation of accretion energy.
We consider the similar slopes as supporting argument to NGC~300~OT being
a scaled-down version of the GE of $\eta$ Car.
{{{}}}The rapid decay in luminosity is determined by the decrease
in the supply of accreted mass.
Later, the slower decay is dictated by the settling of the inner part of the
inflated envelope on the accreting object.
In this later phase the luminosity is below the Eddington luminosity.
The time scale itself depends on the accreting object and on the structure of the inflated envelope.
\begin{figure}[!t]
\resizebox{0.89\textwidth}{!}{\includegraphics{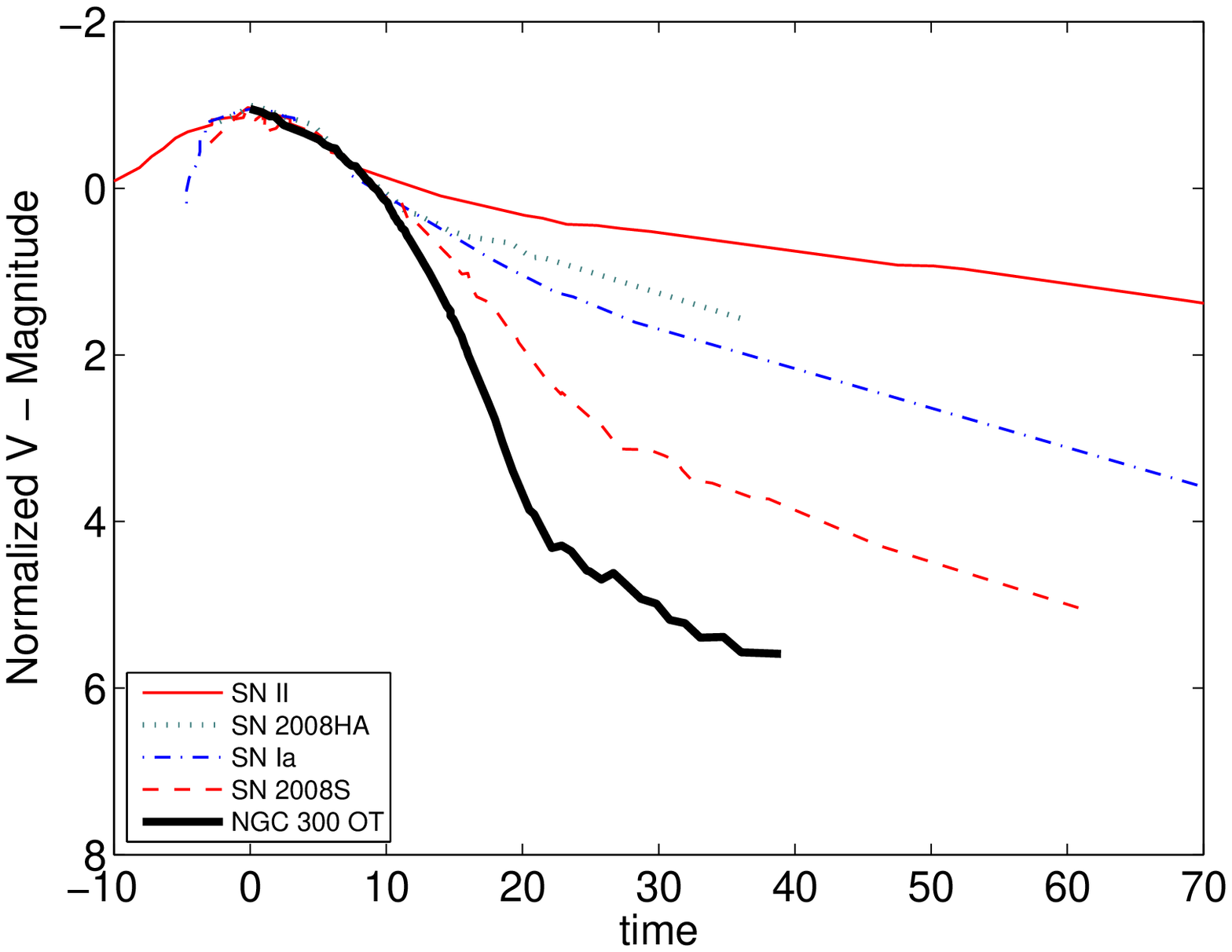}}
\caption{\footnotesize{Comparison of the light curve of NGC~300~OT to SN light curves.
The timescale was normalized so that
1 time unit equals 5.6 days for NGC~300~OT, 4.5 days for SN 2008S,
1.3 days for SN 2008HA,
2.9 days for SN Ia, and 1.3 days for SN II.
The magnitude was shifted to bring peak luminosities to overlap.
It can be easily seen that the decline of NGC~300~OT is much steeper than the decline of SNe.
The references for the data are:
SN 2008S - Botticella et al. (2009); SN Ia (represented by SN 1991T) - Patat et al. (2001);
SN II (represented by SN 1993J) - Patat et al. (2001).
}}
\label{fig:lightcurves2}
\end{figure}

On the other hand, when we compare NGC~300~OT to SNe (Fig. \ref{fig:lightcurves1})
we find that the slope of the decline phase of NGC~300~OT is steeper than
the slope of the decline phase of SNe.
Even SN~2008S which is considered the ``twin'' of NGC~300~OT (Prieto et al. 2008)
shows this discrepancy -- its decay was {{{}}}more gradual than that of
NGC~300~OT in the V-band.
This is despite the fact that its progenitor had properties similar to {{{those}}} of NGC~300~OT,
and the outburst duration was similar.
If SN~2008S is indeed a SN, as argued by Botticella et al. (2009), its kinetic energy
is expected to be $\sim 10-100$ larger than the radiated energy.
In such a case SN~2008S would be located in the region of exploding stars in Figure~\ref{fig:totEvst},
consistent with its claimed SN nature.

To summarize this section, we suggest that the eruption of NGC~300~OT
can be understood as a scaled-down version of the GE of $\eta$ Car,
in terms of a similar dominating physical process $-$ accretion.

\section{THE TOTAL ENERGY BUDGET OF ACCRETION-POWERED ILOTS}
\label{sec:transients}

The objects presented in Section \ref{sec:companion} as examples of the accretion-powered
transient scenario fall between the regions of novae and SNe in the peak visible brightness vs.
event timescale observational diagrams (e.g., Kulkarni \& Kasliwal 2009; Rau et al. 2009).
This region contains more objects that are termed ILOTs.
With the accretion scenario at hand, it may be instructive to plot a diagram based on the
total energy of the transient events instead of their luminosity.

Part of the energy liberated by accretion is not radiated,
but rather channeled to other processes, such as lifting envelope material.
Energy radiated outside the optical bands is usually not observed.
Also, the kinetic energy of the ejecta is hard to estimate due to model-dependent mass loss values,
and a somewhat wide range of measured velocities.
Therefore, there is no unique way to define the total energy of the event.
Nevertheless, we define the total `observed' energy as the bolometric energy
(estimated from the optical measurements) emitted during the outburst, plus the ejecta kinetic energy.
The duration of the outburst is formally defined as the time by which the event brightness
(in the V band) drops by 2mag.
In Figure~\ref{fig:totEvst} this total energy is plotted as a function of the event timescale
for several ILOTs and compared to the values for SNe and classical novae.
Table~\ref{tab:data} presents the data used to plot Figure~\ref{fig:totEvst} with references.

SNe Ib/c, and II are {{{}}} grouped as Exploding Massive Stars (this includes hypernovae).
Notice that for classical novae we plot both the observed range from Della Valle \& Livio (1995)
(the two wavy green lines), and the model values from Yaron et al.(2005) (red dots).
The nova model values include the kinetic energy of the ejecta,
but this is usually much smaller {{{or at most equal}} to {{{}}} the radiated energy (Epelstain et al. 2007).
A few models from Yaron et al. (2005) with extreme, unobserved, values of average velocity ($>2500 \kms$)
and mass loss ($>3 \times 10^{-4} M_\odot$) were not included in the plot.
\begin{table*}
\caption{
Data used to plot the total observed energy vs. event timescale in Figure~\ref{fig:totEvst}.
Total energy included radiative and kinetic energy.
References:
1. Berger et al. (2009),
2. Tylenda \& Soker (2006),
3. Kulkarni et al. (2007a),
4. Kulkarni et al. (2007b),
5. Rau et al. (2007),
6. Pastorello et al. (2007),
7. Ofek et al (2007),
8. Smith et al. (2009),
9. Soker, Frankowski \& Kashi (2009),
10. Barbary et al. (2009)
11. Rich et al. (1989),
12. Mould et al. (1990),
13. Foley et al. (2009),
14. Valenti et al. (2009),
15. Arnett \& Fu (1989),
16. Bethe \& Pizzochero (1990),
17. Davidson \& Humphreys (1997),
18. Smith et al. (2003),19. Frew (2004),
20. Humphreys et al. (1999),
21. Smith (2005),
22. Richardson et al. (2009),
23. Smith \&  Hartigan (2006),
24. Rau et al. (2009),
25. Branch (1992),
26. Richardson et al. (2006),
27. Yaron et al. (2005).
}
\label{tab:data} \medskip
\footnotesize{
\begin{tabular}{p{4cm} l l l p{6cm}} 
\hline \hline
\multicolumn{1}{c}{Name} & \multicolumn{1}{c}{Duration}  & \multicolumn{1}{c}{Total energy}     & \multicolumn{1}{c}{ref.}  & Remarks \\
\multicolumn{1}{c}{}     & \multicolumn{1}{c}{[d]}       & \multicolumn{1}{c}{[$10^{48} \erg$]} & \multicolumn{1}{c}{}      & \multicolumn{1}{c}{}\\
\hline
NGC~300~OT      & $80$    & 0.2--0.4   & (1)         &  \\
V838~Mon        & $90$    & 0.3--1     & (2)         &  \\
M85~2006~OT (SN)& $180$   & 50--100    & (3,4,5,6)
& Assuming it is a weak SN \\
M85~2006~OT (Acc.)& $180$   & 0.1--0.8   & (6,7)
& Assuming it is accretion-powered \\
SN~2008S        & $75$    & $\sim 1.3$  & (8)
&  We added kinetic energy equal to the radiated energy (upper estimate from (7))\\
SCP~06F6        & $150$   & $\sim 50$  & (9,10)
& Assuming the source is extragalactic; we include the rising phase because the profile is symmetric \\
M31~RV          & $70$    & 0.9        & (11,12)     &  \\
SN~2008HA       & $32$    & 2--50      & (13,14)     &  \\
SN 1987A        & $130$   & 1500       & (15,16)     &  \\
$\eta$~Car~GE   & $7300$  & 80         & (17,18)     &  \\
$\eta$~Car~LE   & $1800$  & 3          & (19,20,21,22)
& Lesser Eruption of $\eta$~Car (1890) \\
P~Cygni 1600AD  & $2200$  & 2.5        & (19,20,23)     &  \\
SNe~Ia          & 30--70  & 500--2000  & (24,25)
& Radiated energy is negligible \\
Exploding Massive Stars
                & 20--400 & 100--$10^5$ & (24,26)
& SNe Ib/c, and II including hypernovae\\
Classical Novae & 1--1000& $10^{-4}$--$10^{-2}$  & (3,27)
& Dots are models from (24), the wavy band is computed using luminosity and duration from (3) \\
\hline
\end{tabular}
}
\end{table*}

It is evident that the locus of ILOTs in Figure~\ref{fig:totEvst} is more
separated from that of classical novae than in the luminosity vs. timescale
plane (e.g., Kulkarni \& Kasliwal 2009; Rau et al. 2009).
The difference from SNe is also clearer, even though less striking.
The use of energy instead of luminosity apparently highlights the
physical mechanisms underlying the various transient classes.
The region of the diagram populated by the supposedly
accretion-powered events is tentatively marked with a colored strip.
We take the appearance of this sequence as further evidence
for a common physical mechanism behind these transients (although not
all ILOTs are necessarily powered by accretion).
As for P~Cygni, which is not yet proven to be a binary system, we predict that
a binary companion exists in that system, or that a common envelope occurred in the
recent past.).
\begin{figure}[!t]
\resizebox{0.89\textwidth}{!}{\includegraphics{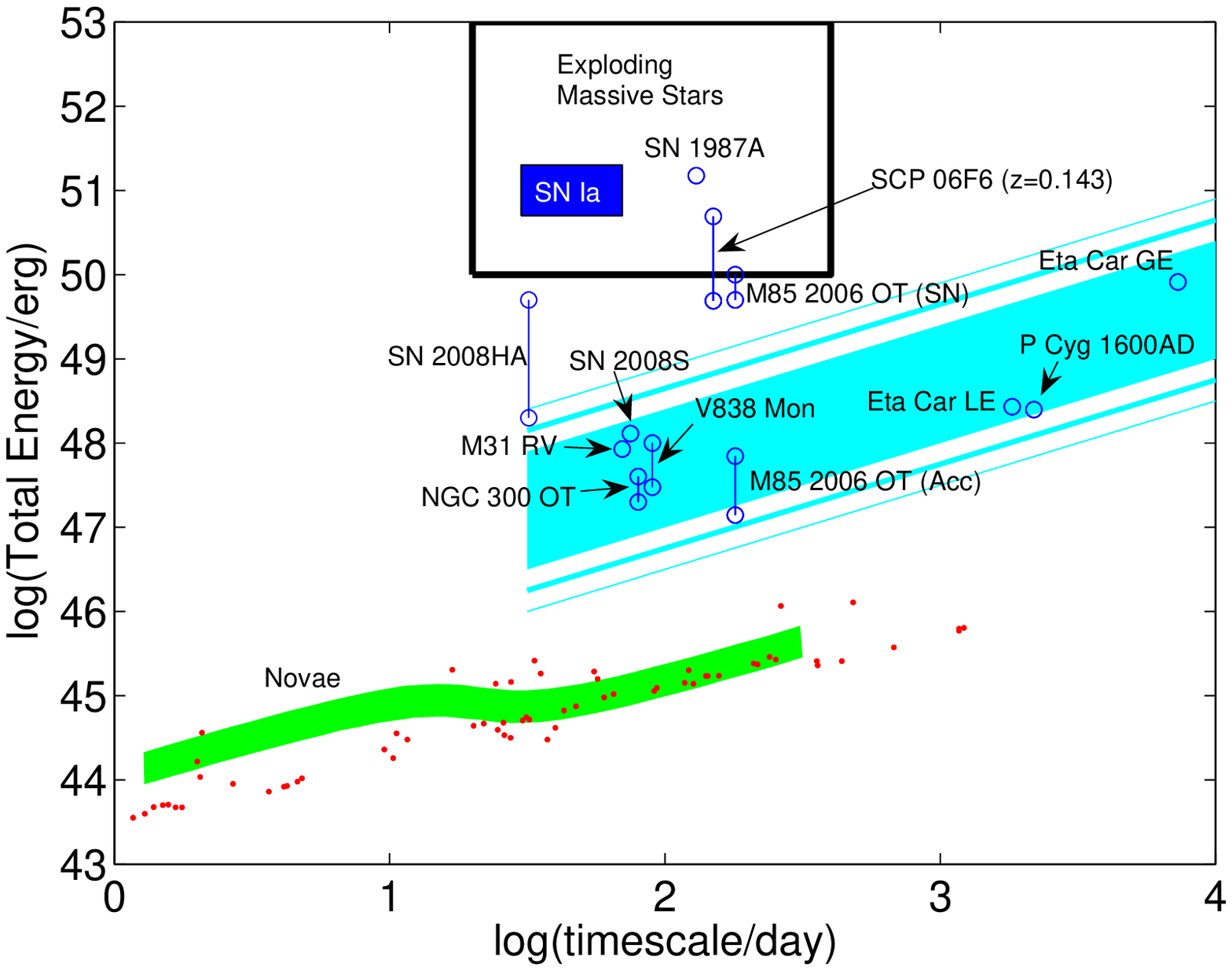}}
\caption{\footnotesize{The total (radiated plus kinetic) energy of the transients discussed in Section
\ref{sec:transients} as a function of the duration of their outbursts.
We did not include the energy which is deposited in lifting the envelope since it is not observed.
See text and Table 1 for information about the different transients and the derivation of their total energy.
The shaded area marks the group of transients we study in the paper.
We propose that most of these transients are associated with mass-transfer processes.}}
\label{fig:totEvst}
\end{figure}

To summarize, we suggest that most (but probably not all) of the
ILOTs are accretion-powered events.
Based on the common energy source they should be grouped in one class
together with stellar merger tidal, disruption flares, and large eruptions in
LBV binary systems.
The shape of the light curve (Fig. \ref{fig:lightcurves1}), related to
the accretion as the energy source, is an important distinguishing feature of
many systems in this group.

We also {{{predict}}} that because of the geometry of
the accretion process the ejecta in this class of transients
should systematically exhibit bipolar structure, as suggested by Soker (2004).
We note that in some single star models circumstellar material can also
posses bipolar structure {{{{(e.g., Smith 2009)}}}}.

We thank Avishay Gal-Yam, Andrea Pastorello, Nathan Smith, Todd Thompson, Romuald Tylenda
and an anonymous referee for helpful comments{{{}}}.
A. F. was supported in part at the Technion by a fellowship from the Lady Davis Foundation.
This research was supported by the Asher Fund for Space Research
at the Technion, and the Israel Science Foundation.


\end{document}